\documentclass[11pt]{article} 
\usepackage{amssymb,amsmath,amsthm}
\usepackage{multicol} 
\usepackage{graphicx} 
\usepackage{amsmath, epsfig} 
\newtheorem{theorem}{Theorem}

\newtheorem{lemma}{Lemma} 

\newtheorem{remark}{Remark} 
\newtheorem{definition}{Definition}

\newcommand{\dalk}{dal(G,k)} 
\newcommand{\damk}{dam(G,k)}
\newcommand{\dk}{dex(G,k)} 
\newcommand{\dmax}{dmax(G)}
\newcommand{\np}{\mathcal{NP}} 
\newcommand{\ignore}[1]{}
\begin{document} 
\title{Finding large and small dense subgraphs} 
\author{Reid Andersen} 
\maketitle

\begin{abstract} We consider two optimization problems related to
    finding dense subgraphs, which are induced subgraphs with high
    average degree.  The densest at-least-$k$-subgraph problem
    (DalkS) is to find an induced subgraph of highest average
    degree among all subgraphs with at least $k$ vertices, and the
    densest at-most-$k$-subgraph problem (DamkS) is defined
    similarly.  These problems are related to the well-known
    densest $k$-subgraph problem (DkS), which is to find the
    densest subgraph on exactly $k$ vertices.  Our main result is
    that DalkS can be approximated efficiently, while DamkS is
    nearly as hard to approximate as the densest $k$-subgraph
    problem.  We give two algorithms for DalkS, a 3-approximation
    algorithm that runs in time $O(m+ n \log n)$, and a
    2-approximation algorithm  that runs in polynomial time.  In
    contrast, we show that if there exists a polynomial time
    approximation algorithm for DamkS with ratio $\gamma$, then
    there is a polynomial time approximation algorithm for DkS with
    ratio $4(\gamma^{2}+\gamma)$.  \end{abstract}

\section{Introduction}

The density of an induced subgraph is the total weight of its edges
divided by the size of its vertex set, or half its average degree. 
The problem of finding the
densest subgraph of a given graph, and various related problems,
have been studied extensively.  In the past decade, identifying
subgraphs with high density has become an important task in the
analysis of large networks \cite{trawling, GKT}.  

There are a variety of efficient algorithms for finding the densest
subgraph of a given graph.  The densest subgraph can be identified
in polynomial time by solving a maximum flow problem
\cite{Goldberg84,GGT89}.  Charikar \cite{Char00} gave a greedy
algorithm that produces a 2-approximation
of the densest subgraph in linear time.  Kannan and Vinay
\cite{Kannan} gave a spectral approximation algorithm for a related
notion of density.  Both of these approximation algorithms are fast
enough to run on extremely large graphs.

In contrast, no practical algorithms are known for finding the
densest subgraph on exactly $k$ vertices.  If $k$ is specified as
part of the input, and is allowed to vary with the graph size $n$,
the best polynomial time algorithm known has approximation ratio
$n^{\delta}$, where $\delta$ is slightly less than $1/3$.  This
algorithm is due to Feige, Peleg, and Korsarz~\cite{FPK01}.  The
densest $k$-subgraph problem is known to be $\np$-complete, but there is a large gap
between this approximation ratio and the strongest known hardness
result.  

In many of the graphs we would like to analyze (for example, graphs
arising from sponsored search auctions, or from links between
blogs), the densest subgraph is extremely small relative to the
size of the graph.  When this is the case, we would like to find a
subgraph that is both large and dense, without solving the
seemingly intractable densest $k$-subgraph problem.  To address
this concern, we introduce the densest at-least-$k$-subgraph
problem, which is to find the densest subgraph on at least $k$
vertices.

In this paper, we show that the densest at-least-$k$-subgraph problem
can be solved nearly as efficiently as the densest subgraph
problem.  In fact, we show it can be solved by a careful
application of the same techniques.  We give a greedy
3-approximation algorithm for DalkS that runs in time $O(m + n \log
n)$ in a weighted graph, and time $O(m)$ in an unweighted graph.
This algorithm is an extension of Charikar's algorithm for densest
subgraph problem.  We also give a 2-approximation algorithm for
DalkS that runs in polynomial time, and can be computed by solving
a single parametric flow problem.  This is an extension of the
algorithm of Gallo, Grigoriadis, and Tarjan \cite{GGT89} for the
densest subgraph problem. 

We also show that finding a dense subgraph with at most $k$
vertices is nearly as hard as finding the densest subgraph with
exactly $k$ vertices.  In particular, we prove that a polynomial
time $\gamma$-approximation algorithm for the densest
at-most-$k$-subgraph problem would imply a polynomial time
$4(\gamma^{2} + \gamma)$-approximation algorithm for the densest
$k$-subgraph problem.  More generally, if there exists a polynomial
time algorithm that approximates DamkS in a weak sense, returning a
set of at most $\beta k$ vertices with density at least $1/\gamma$
times the density of the densest subgraph on at most $k$ vertices,
then there is a polynomial time approximation algorithm for DkS
with ratio $4(\gamma^{2} + \gamma \beta)$.  

Our algorithms for DalkS can find subgraphs with nearly optimal
density in extremely large graphs, while providing considerable
control over the sizes of those subgraphs.  Our reduction of DkS to
DamkS gives additional insight into when DkS is hard, and suggests
a possible approach for improving the approximation ratio for DkS.

The paper is organized as follows.  We first consider the DalkS
problem, presenting the greedy 3-approximation in
Section~\ref{S:atleast}, and the polynomial time 2-approximation in
Section~\ref{S:twoapprox}.  We consider the DamkS problem in
Section~\ref{S:atmost}.
In Section~\ref{S:wishful}, we discuss the possibility of finding
 a good approximation algorithm for DamkS.

\subsection{Related work}

We will briefly survey a few results on the complexity of the
densest $k$-subgraph problem.  The best approximation algorithm known
for the general problem (when $k$ is specified as part of the
input) is the algorithm of Feige, Peleg, and Kortsarz \cite{FPK01},
which has ratio $O(n^{\delta})$ for some $\delta < 1/3$.  For any
particular value of $k$, the greedy algorithm of Asahiro et
al.~\cite{AITT00} gives the ratio $O(n/k)$.  Algorithms based on
linear programming and semidefinite programming have produced
approximation ratios better than $O(n/k)$ for certain values of
$k$, but have not improved the approximation ratio of $n^{\delta}$
for the general case \cite{FS97,FL01}.

Feige and Seltser \cite{FS97} showed the densest $k$-subgraph problem
is $\np$-complete when restricted to bipartite graphs of maximum
degree 3, by a reduction from max-clique.  This reduction does not
produce a hardness of approximation result for DkS.  In fact, they
showed that if a graph contains a $k$-clique, a subgraph with $k$
vertices and $(1-\epsilon) {k \choose 2}$ edges can be found in
subexponential time.  Khot \cite{Khot06} proved there can be no
PTAS for the densest $k$-subgraph problem, under a standard
complexity assumption.

Arora, Karger, and Karpinski~\cite{AKK95} gave a PTAS for the
special case $k=\Omega(n)$ and $m=\Omega(n^{2})$.  Asahiro, Hassin,
and Iwama~\cite{AHI02} showed that the problem is still
$\np$-complete in very sparse graphs.

\section{Definitions}\label{S:defs} Let $G=(V,E)$ be an
undirected graph with a weight function $w : E \rightarrow
\mathbb{R}_{+}$ which assigns a positive weight to each edge.
The weighted degree $w(v,G)$ is the sum of the weights of the
edges incident with $v$.  The total weight $W(G)$ is the sum of
the weights of the edges in $G$.

\begin{definition} For any induced subgraph $H$ of $G$, we define
    the density of $H$ to be $$d(H) = \frac{W(H)}{|H|}.$$
\end{definition} 

\begin{definition} For an undirected graph $G$, we
    define the following quantities.  \begin{align*} \dalk &:=
        \mbox{the maximum density of an induced subgraph on at
        least k vertices.}\\ \damk &:= \mbox{the maximum density of
        an induced subgraph on at most k vertices.}\\ \dk &:=
        \mbox{the maximum density of an induced subgraph on exactly
        k vertices.}\\ \dmax &:= \mbox{the maximum density of any
        induced subgraph.} \end{align*} 
\end{definition}

The densest at-least-$k$-subgraph problem (DalkS) is to find an
induced subgraph on at least $k$ vertices achieving density $\dalk$. 
Similarly, the densest
at-most-$k$-subgraph problem (DamkS) is to find an induced subgraph
on at most $k$ vertices achieving density $\damk$.  
The densest $k$-subgraph problem (DkS) is to find
an induced subgraph on exactly $k$ vertices 
achieving $\dk$, and the densest subgraph
problem is to find an induced subgraph of any size achieving $\dmax$.

We now define formally what it means to be an approximation algorithm for
DalkS.  Approximation algorithms for Damks, DkS, and the densest
subgraph problem, are defined similarly.  \begin{definition} An
    algorithm $A(G,k)$ is a $\gamma$-approximation algorithm for
    the densest at-least-$k$-subgraph problem if for any graph $G$
    and integer $k$, it returns an induced subgraph $H$ on at least
    $k$ vertices of $G$ with density $d(H) \geq \dalk/\gamma$.
\end{definition}

\section{The densest at-least-$k$-subgraph problem}\label{S:atleast}
In this section, we give 3-approximation algorithm for the densest
at-least-$k$-subgraph problem that runs in time $O(m + n \log n)$ in
a weighted graph, and time $O(m)$ in an unweighted graph.  The
algorithm is a simple extension of Charikar's greedy algorithm for
the densest subgraph problem.  To analyze the algorithm, we relate
the density of a graph to the size of its $w$-cores, which are
subgraphs with minimum weighted degree at least $w$.

 \noindent \framebox{ 
 \begin{minipage}{.95\textwidth} 
     {\noindent $ChALK(G,k):$}\\ 
     Input: a graph $G$ with $n$ vertices, and an
     integer $k$.\\ 
     Output: an induced subgraph of $G$ with at
     least $k$ vertices.  
     \begin{enumerate} 
         \item Let $H_{n}=G$ and
             repeat the following step for $i = n, \dots, 1$:
             \begin{enumerate} 
                 \item Let $r_{i}$ be the minimum
                     weighted degree of any vertex in $H_{i}$.
                 \item Let $v_{i}$ be a vertex where
                     $w(v_{i},H_{i}) = r_{i}$.  
                 \item Remove
                     $v_{i}$ from $H_{i}$ to form the induced
                     subgraph $H_{i-1}$.  
             \end{enumerate} \item
                     Compute the density of $d(H_{i})$ for each $i \in
                     [1,n]$.  
                 \item
                     Output the induced subgraph $H_{i}$ maximizing
                     $\max_{i \geq k} d(H_{i})$.  
             \end{enumerate}
         \end{minipage} }

 \begin{theorem}\label{T:Chalk} $ChALK(G,k)$ is a 3-approximation
     algorithm for the densest at-least-$k$-subgraph problem.
 \end{theorem} We will prove Theorem~\ref{T:Chalk} in the following
 subsection.  The implementation of step 1 described by Charikar
  (see \cite{Char00}) gives us the following bound on the running
  time of ChALK.
 \begin{theorem}[Charikar]\label{T:Chalkruntime} The running time
     of $ChALK(G,k)$ is $O(m)$ in an unweighted graph, and $O(m+
     n\log n)$ in a weighted graph.  \end{theorem}

\subsection{Analysis of ChALK} The ChALK algorithm is easy to
understand if we consider the relationship between induced
subgraphs of $G$ with high average degree (dense subgraphs) and
induced subgraphs of $G$ with high minimum degree ($w$-cores).

\begin{definition} Given a graph $G$ and a weight $w \in
    \mathbb{R}$, the {\em $w$-core} $C_{w}(G)$ is the unique
    largest induced subgraph of $G$ with minimum weighted degree at
    least $w$.  \end{definition}

Here is an outline of how we will proceed.  We first prove that the
ChALK algorithm computes all the $w$-cores of $G$
(Lemma~\ref{L:computecores}).  We then prove that for any induced
subgraph $H$ of $G$ with density $d$, the $(2d/3)$-core of $G$ has
total weight at least $W(H)/3$ (Lemma~\ref{L:largecore}).  We will
prove Theorem~\ref{T:Chalk} using these two lemmas.

\begin{lemma}\label{L:computecores} Let $\{H_{1}, \dots, H_{n}\}$,
    $\{v_{1}, \dots, v_{n}\}$, and $\{r_{1}, \dots, r_{n}\}$ be the
    induced subgraphs, vertices, and weighted degrees determined by
    ChALK on the input graph $G$.  For any $w \in \mathbb{R}$, if
    $I(w)$ is the largest index such that $r(v_{I(w)}) \geq w$,
    then $H_{I(w)} = C_{w}(G)$.  \end{lemma} \begin{proof} Fix a
        value of $w$.  It easy to prove by induction that none of
        the vertices $v_{n} \dots v_{I(w)+1}$ that were removed
        before $v_{I(w)}$ is contained in any induced subgraph with
        minimum degree at least $w$.  That implies $C_{w}(G)
        \subseteq H_{I(w)}$.  On the other hand, the minimum degree
        of $H_{I(w)}$ is at least $w$, so $H_{I(w)} \subseteq
        C_{w}(G)$.  Therefore, $H_{I(w)} = C_{w}(G)$.  \end{proof}
        \begin{lemma}\label{L:largecore} For any graph $G$ with
            total weight $W$ and density $d = W/|G|$, the $d$-core
            of $G$ is nonempty.  Furthermore, for any $\alpha \in
            [0,1]$, the total weight of the $(\alpha d)$-core of
            $G$ is strictly greater than $(1-\alpha) W$.
        \end{lemma} \begin{proof} Let $\{H_{1}, \dots, H_{n}\}$ be
            the induced subgraphs  determined by ChALK on the input
            graph $G$.  Fix a value of $w$, let $I(w)$ be the
            largest index such that $r(v_{I(w)}) \geq w$, and
            recall that $H_{I(w)}=C_{w}(G)$ by
            Lemma~\ref{L:computecores}.  Since each edge in $G$ is
            removed exactly once during the course of the
            algorithm, \begin{align*} W &= \sum_{i=1}^{|G|} r(i)\\
                &= \sum_{i=1}^{I(w)} r(i) +
                \sum_{i=I(w)+1}^{|G|}r(i)\\ &< W(H_{I(w)}) + w
                \cdot (|G|-I(w))\\ &\leq W(C_{w}(G)) + w|G|.
            \end{align*} Therefore, \begin{align*} W(C_{w}(G)) > W
                - w|G|.  \end{align*} Taking $w = d = W/|G|$ in the
                equation above, we learn that $W(C_{d}(G)) > 0$.
                Taking $w = \alpha d = \alpha W/|G|$, we learn that
                $W(C_{\alpha d}(G)) > (1-\alpha)W$.

\end{proof}

\begin{proof}[\bf{Proof of Theorem~\ref{T:Chalk}}] Let $\{H_{1},
    \dots, H_{n}\}$ be the induced subgraphs  determined by the
    ChALK algorithm on the input graph $G$.  It suffices to show
    that for any $k$, there is an integer $I \in [k,n]$ satisfying
    $d(H_{I}) \geq \dalk/3$.
    
Let $H_{*}$ be an induced subgraph of $G$ with at least $k$
vertices and with density $d_{*} = W(H_*)/|H_{*}| = \dalk$.  We may
apply Lemma~\ref{L:largecore} to $H_*$ with $\alpha = 2/3$ to show
that $C_{(2d_{*}/3)}(H_{*})$ has total weight at least
$W(H_{*})/3$.  This implies that $C_{(2d_{*}/3)}(G)$ has total
weight at least $W(H_{*})/3$.

The core $C_{(2d_{*}/3)}(G)$ has density at least $d_{*}/3$,
because its minimum degree is at least $2d_{*}/3$.
Lemma~\ref{L:computecores} shows that $C_{(2d_{*}/3)}(G) = H_{I}$,
for $I = |C_{(2d_{*}/3)}(G)|$.  If $I \geq k$, then $H_{I}$
satisfies the requirements of the theorem.  If $I < k$, then
$C_{(2d_{*}/3)}(G) = H_{I}$ is contained in $H_{k}$, and the
following calculation shows that $H_{k}$ satisfies the requirements
of the theorem.  \[d(H_{k}) = \frac{W(H_{k})}{k} \geq
\frac{W(C_{(2d_{*}/3)}(G))}{k} \geq  \frac{W(H_*)/3}{k} =
d_{*}/3.\] \end{proof}

\begin{remark} Charikar proved that $ChALK(G,1)$ is a
    2-approximation algorithm for the densest subgraph problem.
    This can be derived from the fact that if $w = \dmax$, the $w$-core of
    $G$ is nonempty.  \end{remark}

\section{A 2-approximation algorithm for the densest
at-least-$k$-subgraph problem}\label{S:twoapprox} 
In this section, we
will give a polynomial time 2-approximation algorithm for the
densest at-least-$k$ subgraph problem.  The algorithm is based on the
parametric flow algorithm of Gallo, Grigoriadis, and Tarjan
\cite{GGT89}.  It is well-known that the densest subgraph problem
can be solved using similar techniques;  Goldberg \cite{Goldberg84}
showed that the densest subgraph can be found in polynomial time by
solving a sequence of maximum flow problems, and Gallo,
Grigoriadis, and Tarjan described how to find the densest subgraph
using their parametric flow algorithm.  

 It is natural to ask whether there is a polynomial time algorithm
 for the densest at-least-$k$-subgraph problem.  We do not know of
 such an algorithm, nor have we proved that DalkS is
 $\np$-complete.  

\begin{theorem} There is a polynomial time 2-approximation
    algorithm for the densest at-least-$k$-subgraph problem.
\end{theorem}

\begin{proof} The parametric flow algorithm of Gallo, Grigoriadis,
    and Tarjan can compute in polynomial time a collection
    $\mathcal{H}$ of nested induced subgraphs of $G$ such that for
    any value of $\alpha$, the following expression is maximized by
    one of the subgraphs in $\mathcal{H}$.
    \begin{equation}\label{E:ggt} \max_{H \subseteq G} |H| \left(
        d(H) - \alpha \right).    \end{equation}
  
    Let $\mathcal{H}'$ be the modified collection of subgraphs
    obtained by padding each subgraph in $\mathcal{H}$ with
    arbitrary vertices until its size is at least $k$.  We will
    show that there is a set $H \in \mathcal{H'}$ that satisfies
    $d(H)\geq\dalk/2$.  Thus, a polynomial time 2-approximation
    algorithm for DalkS can be obtained by computing $\mathcal{H}$,
    padding some of the sets with arbitrary vertices to form
    $\mathcal{H}'$, and returning the densest set in
    $\mathcal{H'}$.  The running time is dominated by the
    parametric flow algorithm. 

    Let $H_{*}$ be an induced subgraph of $G$ with at least $k$
    vertices that has density $d(H_{*})=\dalk$.  Let $\alpha =
    \dalk/2$, and let $H$ be the set from $\mathcal{H}$ that
    maximizes \eqref{E:ggt} for this value of $\alpha$.  In
    particular,  \begin{equation}\label{ggteqn} |H| (d(H) - \alpha)
        \geq |H_{*}| (d(H_{*}) - \alpha)    \geq |H_{*}|
        d(H_{*})/2.  \end{equation} This implies that $H$ satisfies
        $d(H) \geq \alpha = \dalk/2$.  If $|H| \geq k$, then we are
        done.  If $|H| < k$, then consider the set $H'$ of size
        exactly $k$ obtained by padding $H$ with arbitrary
        vertices.  We will show that $d(H') \geq \dalk/2$, which
        will complete the proof.  First, notice that \eqref{ggteqn}
        implies a lower bound on the size of $H$.
        \begin{equation*} |H| \geq |H_{*}| \frac{d(H_{*})}{2d(H)} =
            |H_{*}| \frac{\dalk}{2d(H)}.  \end{equation*} We can
            then bound the density of the padded set $H'$.
            \begin{align*} d(H') &\geq d(H)
                \left(\frac{|H|}{k}\right) \\ &\geq d(H)
                \left(\frac{|H_{*}|}{k} \frac{\dalk}{2d(H)}
                \right)\\ &= \frac{\dalk}{2}\frac{|H_{*}| }{k} \\
                &\geq \frac{\dalk}{2}.\\ \end{align*} \end{proof}

\section{The densest at-most-$k$-subgraph problem}\label{S:atmost} In
this section, we show that the densest at-most-$k$-subgraph problem
is nearly as hard to approximate as the densest $k$-subgraph
problem.  We will show that if there exists a polynomial time
algorithm that approximates DamkS in a weak sense, returning a set
of at most $\beta k$ vertices with density at least $1/\gamma$
times the density of the densest subgraph on at most $k$ vertices,
then there exists a polynomial time approximation algorithm for DkS
with ratio $4(\gamma^{2} + \gamma \beta)$.  As an immediate
consequence, a 
polynomial time $\gamma$-approximation algorithm for the densest
at-most-$k$-subgraph problem would imply a polynomial time
$4(\gamma^{2} + \gamma)$-approximation algorithm for the densest
$k$-subgraph problem.

\begin{definition} An algorithm $A(G,k)$ is a {\em
    $(\beta,\gamma)$-algorithm} for the densest at-most-$k$-subgraph
    problem if for any input graph $G$ and integer $k$, it returns
    an induced subgraph of $G$ with at most $\beta k$ vertices and
    density at least $\damk/\gamma$.  \end{definition}

\begin{theorem}\label{T:reduction} If there is a polynomial time
    $(\beta,\gamma)$-algorithm for the densest at-most-$k$-subgraph
    problem (where $\beta$ and $\gamma$ are at least 1), then there
    is a polynomial time $4(\gamma^{2} + \gamma
    \beta)$-approximation algorithm for the densest $k$-subgraph
    problem.  \end{theorem}

\begin{proof} Assume there exists a polynomial time algorithm
    $A(G,k)$ that is $(\beta,\gamma)$-algorithm for DamkS.  We will
    now describe a polynomial time approximation algorithm for DkS
    with ratio $4(\gamma^{2} + \gamma \beta)$.

    Given as input a graph $G$ and integer $k$, let $H_{1}=G$, let
    $i=1$, and repeat the following procedure.  Let $H_{i} =
    A(G_{i},k)$ be an induced subgraph of $G_{i}$ with at most
    $\beta k$ vertices and with density at least
    $dam(G_{i},k)/\gamma$.  Remove all the edges in $H_{i}$ from
    $G_{i}$ to form a new graph $G_{i+1}$ on the same vertex set as
    $G$.  Repeat this procedure until all edges have been removed
    from $G$.

    Let $n_{i}$ be the number of vertices in $H_{i}$, let $W_{i} =
    W(H_{i})$, and let $d_{i} = d(H_{i}) = W_{i}/n_{i}$.  Let
    $H_{*}$ be an induced subgraph of $G$ with exactly $k$ vertices
    and density $d_{*}=\dk$.  Notice that if $(W_{1} + \dots +
    W_{t-1}) \leq W(H_{*})/2$, then $d_{t} \geq d_{*}/2\gamma$.
    This is because $d_{t}$ is at least $1/\gamma$ times the
    density of the induced subgraph of $G_{t}$ on the vertex set of
    $H_*$, which is at least \[ \frac{W(H_{*}) - (W_{1} + \dots +
    W_{t-1})}{k} \geq \frac{W(H_{*})}{2k} = \frac{d_{*}}{2} . \]

Let $T$ be the smallest integer such that $(W_{1} + \dots + W_{T})
\geq W(H_{*})/2$, and let $U_{T}$ be the induced subgraph on the
union of the vertex sets of $H_{1}, \dots, H_{T}$.  The total
weight $W(U_{T})$ is at least $W(H_{*})/2$.  The density of $U_{T}$
is \[d(U_{T}) = \frac{W(U_{T})}{|U_{T}|} \geq \frac{W_{1} + \dots +
W_{T}}{n_{1} + \dots + n_{T}} \geq \min_{1 \leq t \leq
T}\frac{W_{t}}{n_{t}} \geq \frac{d_{*}}{2\gamma}.\] To bound the
number of vertices in $U_{T}$, notice that $(n_{1} + \dots +
n_{T-1}) \leq \gamma k$, because 
\[ 
\frac{d_{*}k}{2} 
= \frac{W(H_{*})}{2} 
\geq
\sum_{i=1}^{T-1}W_{i}
=
\sum_{i=1}^{T-1}n_{i}d_{i} \geq \frac{d_{*}}{2\gamma}
\sum_{i=1}^{T-1}n_{i} .\] 

Since $n_{T}$ is at most $\beta k$, we
have $|U_{T}| \leq (n_{1} + \dots + n_{T}) \leq (\gamma+\beta)k$.

There are now two cases to consider.  If $|U_{T}| \leq k$, we add
vertices to $U_{T}$ arbitrarily to form a set $U_{T}'$ of size
exactly $k$.  The set $U_{T}'$ is more than dense enough to prove
the theorem, \[d(U_{T}') \geq \frac{W(H_{*})/2}{k} =
\frac{d_{*}}{2}.\] If $|U_{T}| > k$, then we employ a simple greedy
procedure to reduce the number of vertices.  We begin with the
induced subgraph $U_{T}$, greedily remove the vertex with smallest
degree to obtain a smaller subgraph, and repeat until exactly $k$
vertices remain.  The resulting subgraph $U_{T}''$ has density at
least $d(U_{T})(k/2|U_{T}|)$ by the method of conditional
expectations (see also \cite{FPK01}).  The set $U_{T}''$ is
sufficiently dense, \[ d(U_{T}'') \geq d(U_{T}) \frac{k}{2|U_{T}|}
\geq \left(\frac{d_{*}}{2\gamma}\right)
\left(\frac{k}{2(\gamma+\beta)k}\right) = \frac{d_{*}}{4(\gamma^{2}
+ \gamma \beta)} .  \] \end{proof}

\begin{remark} The argument from Theorem~\ref{T:reduction} proves a
    slightly more general statement: if there is a polynomial time
    algorithm for DamkS that is a $(\beta,\gamma)$-algorithm 
    for certain values of $k$, then there is a polynomial time
    algorithm for DkS that is a 
    $4(\gamma^{2} + \gamma \beta)$-approximation algorithm 
    for those same values of  $k$.  \end{remark}
    
We remark that the densest at-most-$k$-subgraph is easily seen to be
$\np$-complete, since a subgraph of size at most $k$ has density at
least $(k-1)/2$ if and only if it is a $k$-clique.  As mentioned
previously, Feige and Seltser~\cite{FS97} proved that the densest
$k$-subgraph problem remains $\np$-complete when restricted to graphs
with maximum degree 3, and their proof shows that the same
statement is true for the densest at-most-$k$-subgraph problem.

\section{Conclusion}\label{S:wishful}

In this section, we discuss the possibility of improving the
approximation ratio for DkS via an approximation algorithm for
DamkS.  One possible approach is to develop a {\em local algorithm}
for DamkS, analogous to the recently developed local algorithms for
graph partitioning~\cite{Spielman,Andersen}. 
For any partition separating $k$ vertices, 
these algorithms can produce a partition separating $O(k)$
vertices that is nearly as good (in terms of conductance).

We conjecture that there is a local algorithm for the densest
subgraph problem that finds a subgraph of density at least
$\theta/\log n$ on at most $O(k^{1+\delta})$ vertices, whenever
there exists a subgraph of density $\theta$ on $k$ vertices.  This
would be a $(\log n,k^{\delta})$-approximation algorithm for DamkS,
which would lead to an approximation algorithm for the densest
$k$-subgraph problem with ratio $O(k^{\delta} \log^{2}n)$.  An
algorithm with $\delta=1$ would not be helpful for approximating
DkS, since an approximation ratio of $O(k)$ can be obtained
trivially.  At the other extreme, an algorithm with $\delta=0$
would produce an $O( \log^{2} n)$ approximation algorithm for DkS,
which seems unlikely.

\begin{small}

\end{small}


\begin{thebibliography}{77}	

\bibitem{Andersen} R.~Andersen, F.~Chung, and K.~Lang.  \newblock
    Local graph partitioning using PageRank vectors.  \newblock In
    {\it Proc. 47th Annual Symposium on Foundations of Computer
    Science (FOCS 2006)}, pp. 475--486.  

\bibitem{AKK95} S. Arora, D. Karger and M. Karpinski. Polynomial
    time approximation schemes for dense instances of NP-hard
    problems. In {\it Proc.  27th ACM Symposium on Theory of
    Computing (STOC 1995)} pp. 284--293.

\bibitem{AHI02} Y. Asahiro, R. Hassin and K. Iwama, Complexity of
    finding dense subgraphs, {\it Discrete Appl. Math.} 121(1-3),
    pp. 15--26, 2002.

\bibitem{AITT00} Y. Asahiro, K. Iwama, H. Tamaki and T. Tokuyama,
    Greedily finding a dense subgraph, {\it J. Algorithms} 34(2),
    pp. 203--221, 2000. 

\bibitem{Char00} M.~Charikar, Greedy approximation algorithms for
    finding dense components in a graph, Proceedings Third
    International Workshop on Approximation Algorithms for
    Combinatorial Optimization, {\it Lecture Notes in Computer
    Science}
    vol. 1913, Springer, Berlin, pp. 84--95, 2000.

\bibitem{FL01} U.~Feige and M.~Langberg.  Approximation algorithms
    for maximization problems arising in graph partitioning.  {\it
    J. Algorithms} 41(2), pp. 174--211, 2001.

\bibitem{FPK01} U.~Feige, D.~Peleg, and G.~Kortsarz.  The dense
    k-subgraph problem.  {\it Algorithmica}, 29(3), 410--421,
    2001.

\bibitem{FS97} U.~Feige and M.~Seltser, On the densest k-subgraph
    problem, Technical report, Department of Applied Mathematics
    and Computer Science, The Weizmann Institute, Rehobot, 1997.

\bibitem{GGT89} G.~Gallo, M.~Grigoriadis and R.~Tarjan, A fast
    parametric maximum flow algorithm and applications, 
    {\it SIAM J.  Comput.} 18(1), pp. 30-55, 1989. 

\bibitem{GKT} D.~Gibson, R.~Kumar, and A.~Tomkins.  Discovering
    large dense subgraphs in massive graphs.  In {\it Proc. 31st
    VLDB Conference}, 2005. 

\bibitem{Goldberg84} A.~Goldberg, Finding a maximum density
    subgraph, Technical Report UCB/CSB 84/171, Department of
    Electrical Engineering and Computer Science, University of
    California, Berkeley, CA, 1984.

\bibitem{Kannan} R.~Kannan and V.~Vinay.  Analyzing the structure
    of large graphs.  {\it Manuscript}, 1999.

\bibitem{Khot06} S.~Khot. Ruling out PTAS for graph min-bisection,
    dense k-subgraph, and bipartite clique, 
    {\it SIAM Journal on Computing}, 36(4), pp. 1025--1071, 2006.

\bibitem{trawling} R.~Kumar, P.~Raghavan, S.~Rajagopalan, and
    A.~Tomkins.  
    Trawling the Web for emerging cyber-communities.
    In {\it Proc. 8th WWW Conference (WWW 1999)}. 

\bibitem{Spielman} D.~Spielman and S.H.~Teng.  Nearly-linear time
    algorithms for graph partitioning, graph sparsification, and
    solving linear systems.  In {\it Proc. 36th Annual ACM
    Symposium on Theory of Computing (STOC 2004)}.


\end{thebibliography}
\end{document}